# Interaction effects in assembly of magnetic nanoparticles

N. A. Usov[1,2], O. N. Serebryakova[1,2] and V. P. Tarasov[1]

[1]*National University of Science and Technology «MISIS», 119049, Moscow, Russia*
[2]*Pushkov Institute of Terrestrial Magnetism, Ionosphere and Radio Wave Propagation, Russian Academy of Sciences, (IZMIRAN) 108480, Troitsk, Moscow, Russia*

**Abstract**
A specific absorption rate of a dilute assembly of various random clusters of iron oxide nanoparticles in alternating magnetic field has been calculated using Landau- Lifshitz stochastic equation. This approach simultaneously takes into account both the presence of thermal fluctuations of the nanoparticle magnetic moments, and magneto-dipole interaction between the nanoparticles of the clusters. It is shown that for usual 3D clusters the intensity of magneto-dipole interaction is determined mainly by the cluster packing density $\eta = N_p V/V_{cl}$, where $N_p$ is the average number of the particles in the cluster, $V$ is the nanoparticle volume, and $V_{cl}$ is the cluster volume. The area of the low frequency hysteresis loop and the assembly specific absorption rate have been found to be considerably reduced when the packing density of the clusters increases in the range of $0.005 \leq \eta < 0.4$. The dependence of the specific absorption rate on the mean nanoparticle diameter is retained with increase of $\eta$, but becomes less pronounced. For fractal clusters of nanoparticles, which arise in biological media, in addition to considerable reduction of the absorption rate, the absorption maximum is shifted to smaller particle diameters. It is found also that the specific absorption rate of fractal clusters increases appreciably with increase of the thickness of nonmagnetic shells at the nanoparticle surfaces.

**Introduction**

Magnetic hyperthermia [1-4] is one of the most promising directions in contemporary biomedical research related with the cancer treatment. The performance of magnetic nanoparticles to generate heat in alternating external magnetic field is affected by various factors, such as their geometrical and material parameters, the concentration of nanoparticles in the media, as well as the frequency and amplitude of the alternating magnetic field. In this paper the effect of mutual magneto-dipole interaction on the specific absorption rate (SAR) of an assembly of magnetic nanoparticles an alternating magnetic field is studied theoretically. The nanoparticles of iron oxides seem most promising for use in magnetic hyperthermia [2-5], because they are biocompatible and biodegradable, and can be detected in human body using clinical MRI. In this study we consider assemblies of nanoparticles with magnetic parameters typical of iron oxide nanoparticles. It has been found recently [4,6] that being embedded in a biological environment, for example, into a tumor, magnetic nanoparticles turn out to be tightly bound to the surrounding tissues. Therefore, the rotation of magnetic nanoparticles as a whole under the influence of alternating external magnetic field is greatly hindered. In such a case, the Brownian relaxation is unimportant [4]. Therefore, only the motion of the particle magnetic moments under the influence of an alternating magnetic field and thermal fluctuations has to be considered. In addition, one must take into account the influence of the magnetic-dipole interaction between particles. The latter effect is especially important since magnetic nanoparticles in biological media tend to agglomerate [2,4,7] forming dense aggregates of nanoparticles having fractal [8,9] geometrical structure.

The effect of thermal fluctuations on the heat dissipation in a dilute assembly of magnetic nanoparticles in alternating magnetic field has been studied in detail in Refs. 10-13. In particular, it has been shown [10] that the SAR of such assembly depends substantially on the mean nanoparticle diameter, among other factors. For a dilute nanoparticle assembly detailed calculations [10] allow one to determine the optimum diameter of the nanoparticles at the given particle magnetic parameters and given amplitude and frequency of the alternating magnetic field. With optimal choice of geometric and magnetic parameters of the nanoparticles very high SAR values, of the order of 1 kW/g, have been predicted [10,11]. It is notable that the SAR values reported in a number of experiments [14-17] are really close to the above theoretical estimates. At the same time, in many experiments [5,18-21] significantly lower values of SAR ~ 20 - 50 W/g were measured. This fact can be explained, most likely, by the influence of strong magneto-dipole interaction in dense assemblies of magnetic nanoparticles.

Indeed, it has been experimentally shown [22,23] that the SAR in the dense assembly of magnetic nanoparticles essentially depends on the aspect ratio of the test sample, i.e. the ratio of sample length to width. This is indirect evidence of the influence of magneto-dipole interaction on the response of an assembly of nanoparticles on alternating external magnetic field. The effect of magneto-dipole interaction on the energy absorption rate by the assembly of magnetic nanoparticles has been studied in a number of recent theoretical and experimental investigations [7,24-38]. However, further investigations seem necessary to take into account fractal nature [8,9] of the nanoparticle distribution in biological media.

To see clearly the effect of magneto-dipole interaction, in this paper we first calculate the SAR of an assembly of non-interacting iron oxide nanoparticles. To study the effect of magneto-dipole interaction, we solve numerically the Landau- Lifshitz stochastic equation [13,39-41], which simultaneously takes into account both the presence of thermal fluctuations of the particle magnetic moments, and magneto-dipole interaction between the nanoparticles of the clusters. Two types of magnetic clusters are considered, the usual random 3D clusters of nanoparticles distributed in a rigid media and fractal clusters of nanoparticles with usually arise within the intracellular space. Note that within the cluster the nanoparticles are coupled by a strong magneto- dipole interaction. At the same time, for a dilute assembly of clusters the magnetic interaction between the clusters can be neglected in a first approximation.

The influence of magneto- dipole interaction on the properties of a dilute assembly of random 3D clusters is shown to be determined mainly by the nanoparticle packing density $\eta = N_p V/V_{cl}$, where $N_p$ is the average number of particles in the cluster, $V$ is the nanoparticle volume, and $V_{cl}$ is the cluster volume.. The area of the hysteresis loop, and the assembly SAR are found to be considerably reduced when the packing density of the 3D clusters increases in the range of packing densities studied, $0.005 \leq \eta < 0.4$. For fractal clusters of magnetic nanoparticles, in addition to considerable reduction of SAR, the maximum of absorption rate is shifted to smaller particle diameters, as a rule. It is found also that the SAR of fractal clusters increases appreciably with increase of the thickness of nonmagnetic shells at the nanoparticle surfaces. This effect may be important for application of magnetic nanoparticle assemblies in magnetic hyperthermia.

**Numerical simulation**
**Non interacting nanoparticles**

It is instructive to remind first the behavior of an assembly of non interacting superparamagnetic nanoparticles in an alternating magnetic field. It enables one to see clearly the influence of the magneto-dipole interaction on the assembly properties. Based on the Fokker-Planck equation derived by W.F. Brown [39], one can get an approximate kinetic equation [10] for the population numbers $n_1(t)$ and $n_2(t)$ of two potential wells of uniaxial superparamagnetic nanoparticle

$$\frac{\partial n_1}{\partial t} = \frac{n_2}{\tau_2(T)} - \frac{n_1}{\tau_1(T)}; \quad n_1(t) + n_2(t) = 1 \quad (1)$$

Here $\tau_1(T)$ and $\tau_2(T)$ are the corresponding relaxation times at a given temperature $T$ for first and second potential wells, respectively. The relaxation times $\tau_1(T)$ and $\tau_2(T)$ depend essentially on the amplitude and direction of the applied magnetic field with respect to particle easy anisotropy axis (see Appendix in Ref. 10).

The iteration procedure can be used to calculate the well population numbers $n_1(t)$ and $n_2(t)$ for several periods of the alternating magnetic field. It is sufficient to obtain stationary hysteresis loop of a particle in alternating magnetic field. To do so one can use an approximate relation for the component of the reduced particle magnetization along the magnetic field direction

$$\frac{M_h}{M_s V} = m_h(t) = n_2(t)\cos[\theta_0 - \theta_{\min,2}(h_e(t))] + n_1(t)\cos[\theta_0 - \theta_{\min,1}(h_e(t))] \quad (2)$$

Here $\theta_0$ is the angle of the external magnetic field with respect to the particle easy anisotropy axis, $\theta_{\min,1}$ and $\theta_{\min,2}$ are the locations of the potential well minima as the functions of the reduced applied magnetic field, $h_e(t) = H_0 \sin(\omega t)/H_a$, where $\omega = 2\pi f$ is the angular frequency, $H_a$ being the particle anisotropy field. To get hysteresis loop of an assembly of randomly oriented independent nanoparticles it is necessary to average the reduced magnetization $m_h(t)$ over the magnetic field directions.

It is worth noting that the accuracy of an approximate analytical solution, Eq. (1), (2), of the Fokker-Planck equation has been validated [10] by means of direct comparison with the numerical solutions of the stochastic Landau-Lifshitz equation for non interacting magnetic nanoparticles.

**Nanoparticle clusters**

To investigate the effect of magneto-dipole interaction on the specific absorption rate of an assembly of interacting magnetic nanoparticles in an alternating magnetic field in this paper we study the behavior of a dilute assembly of usual 3D clusters of superparamagnetic nanoparticles and that of fractal clusters [8, 9] which arise usually in biological media loaded with fine magnetic nanoparticles.

A quasi-spherical 3D cluster of nanoparticles shown schematically in Fig. 1a can be characterized by its radius $R_{cl}$, and the number of nanoparticles, $N_p \gg 1$, within its volume. It is assumed that the nanoparticles have nearly the same diameter $D$, and their centers, $\{r_i\}$, $i = 1,2, N_p$, are randomly distributed in the cluster volume. We also assume that the particles are coated with thin non-magnetic shells, so that the exchange interaction between the neighboring nanoparticles of the cluster is absent. As we mentioned above, such 3D cluster is characterized by the nanoparticle packing density $\eta = N_p V/V_{cl}$. This is a total volume of the magnetic material distributed in the volume of the cluster. One can define the average distance between the nanoparticles of the cluster by means of the relation $D_{av} = (6V_{cl}/\pi N_p)^{1/3}$. Then, the nanoparticle packing density is given by $\eta = (D/D_{av})^3$.

For an assembly of completely random 3D clusters the orientations of the easy anisotropy axes of nanoparticles $\{e_i\}$, $i = 1,2, N_p$, are chosen randomly and independently on the unit sphere. Alternatively, one can assume that during the formation of clusters in a solution under the influence of magneto-static interaction certain correlation occurs in the distribution of the nanoparticle easy anisotropy axis directions. One possibility to describe such partially ordered clusters is to assume that the easy anisotropy axes of the nanoparticles are uniformly distributed in a solid angle, $\theta < \theta_{max}$, in the spherical coordinates.

Random 3D clusters with a given number of particles $N_p$ of diameter $D$ were created in this study as follows. First, we generated dense enough and approximately uniform set of $N$ random points $\{\boldsymbol{\rho}_i\}$ within a spherical volume of the radius $R_{cl}$, so that $|\boldsymbol{\rho}_i| \leq R_{cl}$ for all generated points, $i = 1, 2, ...N$, $N >> N_p$. Center of the first nanoparticle was placed in the first random point, $\boldsymbol{r}_1 = \boldsymbol{\rho}_1$. Then all random points with coordinates $|\boldsymbol{\rho}_i - \boldsymbol{r}_1| \leq D$ were removed from the initial set of the random points. After this operation, any point in the remaining set of random points could be used as a center of the second nanoparticle. For example, one can put simply $\boldsymbol{r}_2 = \boldsymbol{\rho}_2$. At the next step one removes all random points whose coordinates satisfy the inequality $|\boldsymbol{\rho}_i - \boldsymbol{r}_2| \leq D$. This procedure is repeated until all $N_p$ nanoparticle centers are placed within the cluster volume. As a result, all random nanoparticle centers lie within a sphere of radius $R_{cl}$, so that $|\boldsymbol{r}_i| \leq R_{cl}$, $i = 1, 2, .. N_p$. Furthermore, none of the nanoparticles is in a direct contact with the neighboring nanoparticles. This algorithm enables one to construct random quasi-spherical 3D clusters of magnetic nanoparticles for moderate values of the nanoparticle volume fraction $\eta < 0.5$.

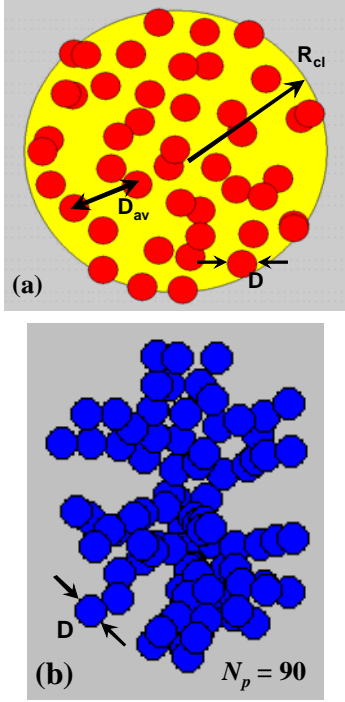

Fig. 1. Geometry of quasi-spherical random 3D cluster of single-domain nanoparticles (a) and fractal cluster (b) with fractal descriptors $D_f = 2.1$ and $k_f = 1.3$.

For a given set of initial parameters, i.e. $D$, $R_{cl}$ and $N_p$, various random 3D clusters differ by the sets of the coordinates of the nanoparticle centers $\{\boldsymbol{r}_i\}$, and orientations $\{\boldsymbol{e}_i\}$ of the particle easy anisotropy axes. However, the calculations show that in the limit $N_p >> 1$ the hysteresis loops obtained for different realizations of random variables $\{\boldsymbol{r}_i\}$ and $\{\boldsymbol{e}_i\}$ differ only slightly from each other. To characterize the behavior of a dilute assembly of random nanoparticle clusters it is necessary to calculate assembly hysteresis loop averaged over a sufficiently large number of random cluster realizations. The latter characterizes the behavior of a dilute assembly of random nanoparticle clusters. It is found that in the limit $N_p >> 1$ the averaged hysteresis loop of cluster assembly has a rather small dispersion even being averaged over 20 - 30 independent realizations of random clusters with the fixed values of the initial parameters $D$, $R_{cl}$ and $N_p$.

The geometry of fractal clusters of single-domain nanoparticles is characterized [42, 43] by the fractal descriptors $D_f$ and $k_f$. By definition, the total number of nanoparticles $N_p$ in the fractal cluster is given by the relation $N_p = k_f (2R_g/D)^{D_f}$, where $D_f$ is the fractal dimension, $k_f$ is the fractal prefactor, $R_g$ being the radius of gyration. It is defined [43] via the mean square of the distances between the particle centers and the geometrical center of mass of the aggregate. In this paper the fractal clusters with various fractal descriptors were created using the well known Filippov's et. al. algorithm [43]. As an example, Fig. 1b shows the geometrical structure of fractal cluster with fractal descriptors $D_f = 2.1$ and $k_f = 1.3$ consisting of $N_p = 90$ single-domain nanoparticles. Geometrically, it seems that the main difference between 3D and fractal clusters is that in the latter case every nanoparticle has at least one neighbor located at closest possible distance between nanoparticle centers equal to the nanoparticle diameter $D$.

Dynamics of unit magnetization vector $\vec{\alpha}_i$ of $i$-th single-domain nanoparticle of the cluster is determined by the stochastic Landau-Lifshitz (LL) equation

$$\frac{\partial \vec{\alpha}_i}{\partial t} = -\gamma_1 \vec{\alpha}_i \times \left( \vec{H}_{ef,i} + \vec{H}_{th,i} \right) - \kappa \gamma_1 \vec{\alpha}_i \times \left( \vec{\alpha}_i \times \left( \vec{H}_{ef,i} + \vec{H}_{th,i} \right) \right) \quad (3)$$

where $\gamma$ is the gyromagnetic ratio, $\kappa$ is phenomenological damping parameter, $\gamma_1 = \gamma/(1+\kappa^2)$, $\vec{H}_{ef,i}$ is the effective magnetic field and $\vec{H}_{th,i}$ is the thermal field. The effective magnetic field acting on a separate nanoparticle can be calculated as a derivative of the total cluster energy

$$\vec{H}_{ef,i} = -\frac{\partial W}{VM_s \partial \vec{\alpha}_i} \quad (4)$$

The total magnetic energy of the cluster $W = W_a + W_Z + W_m$ is a sum of the magneto-crystalline anisotropy energy $W_a$, Zeeman energy $W_Z$ of the particles in applied magnetic field $\vec{H}_0 \sin(\omega t)$, and the energy of mutual magneto-dipole interaction of the particles $W_m$.

For nanoparticles of nearly spherical shape with uniaxial type of magnetic anisotropy, the magneto-anisotropy energy is given by

$$W_a = KV \sum_{i=1}^{N_p} \left(1 - (\vec{\alpha}_i \vec{e}_i)^2\right) \quad (5)$$

where $e_i$ is the orientation of the easy anisotropy axis of $i$-th particle of the cluster. Zeeman energy $W_Z$ of the cluster in applied magnetic field is given by

$$W_Z = -M_s V \sum_{i=1}^{N_p} \left(\vec{\alpha}_i \vec{H}_0 \sin(\omega t)\right) \quad (6)$$

Next, for spherical uniformly magnetized nanoparticles the magnetostatic energy of the cluster can be represented as the energy of the point interacting dipoles located at the particle centers $r_i$ within the cluster. Then the magneto-dipole interacting energy is

$$W_m = \frac{M_s^2 V^2}{2} \sum_{i \neq j} \frac{\vec{\alpha}_i \vec{\alpha}_j - 3(\vec{\alpha}_i \vec{n}_{ij})(\vec{\alpha}_j \vec{n}_{ij})}{|\vec{r}_i - \vec{r}_j|^3} \quad (7)$$

where $n_{ij}$ is the unit vector along the line connecting the centers of $i$-th and $j$-th particles, respectively.

Thus, the effective magnetic field acting on the $i$-th nanoparticle of the cluster is given by

$$\vec{H}_{ef,i} = H_a (\vec{\alpha}_i \vec{e}_i) \vec{e}_i + \vec{H}_0 \sin(\omega t) + M_s V \sum_{j \neq i} \frac{\vec{\alpha}_j - 3(\vec{\alpha}_j \vec{n}_{ij}) \vec{n}_{ij}}{|\vec{r}_i - \vec{r}_j|^3} \quad (8)$$

where $H_a = 2K/M_s$ is the particle anisotropy field.

The thermal fields, $\vec{H}_{th,i}$, $i = 1, 2, ..N_p$, acting on various nanoparticles of the cluster are statistically independent, with the following statistical properties [39] of their components for every nanoparticle

$$\langle H_{th}^{(\alpha)}(t) \rangle = 0;$$
$$\langle H_{th}^{(\alpha)}(t) H_{th}^{(\beta)}(t_1) \rangle = \frac{2k_B T \kappa}{\gamma M_s V} \delta_{\alpha\beta} \delta(t - t_1);$$
$$\alpha, \beta = (x, y, z). \quad (9)$$

Here $k_B$ is the Boltzmann constant, $\delta_{\alpha\beta}$ is the Kroneker symbol, and $\delta(t)$ is the delta function.

The procedure for solving stochastic differential equation (3), (8) and (9) is described in detail in Refs. 13, 40 and 41.

**Results and Discussion**
**Non interacting iron oxide nanoparticles**

Consider a dilute assembly of superparamagnetic nanoparticles with an average diameter $D$. The particles are assumed to be tightly packed in a surrounding media, and their easy anisotropy axes are randomly oriented in space. The hysteresis loop of such an assembly in alternating magnetic field $H = H_0 \sin(\omega t)$ can be calculated [10] using equations (1), (2). This approach, due to its simplicity, allows one to carry out detailed calculations of the assembly hysteresis loops for various particle sizes depending on frequency and amplitude of the alternating magnetic field. In the calculations performed, in accordance with the experimental data [2-6], the saturation magnetization of iron oxide nanoparticles is assumed to be $M_s = 350$ emu/cm$^3$, the magnetic anisotropy constant being $K = 10^5$ erg/cm$^3$. The assembly temperature is $T = 300$ K, the nanoparticle diameters are in the range $D = 10 - 30$ nm. These parameters seem typical for experiments carried out on iron oxide nanoparticles.

Fig. 2 shows the SAR of non-interacting assemblies of iron oxide nanoparticles at various frequencies at a fixed amplitude of alternating magnetic field, $H_0 = 100$ Oe. As can be seen, for the range of frequencies that are characteristic for magnetic hyperthermia, $f = 200 - 500$ kHz, SAR has a maximum for the assembly of iron oxide nanoparticles with diameters $D = 20 - 21$ nm. It is notable that even at relatively moderate amplitude of alternating magnetic field the assembly SAR reaches sufficiently high values, 0.35 - 0.45 kW/g, if the nanoparticle diameters are chosen properly.

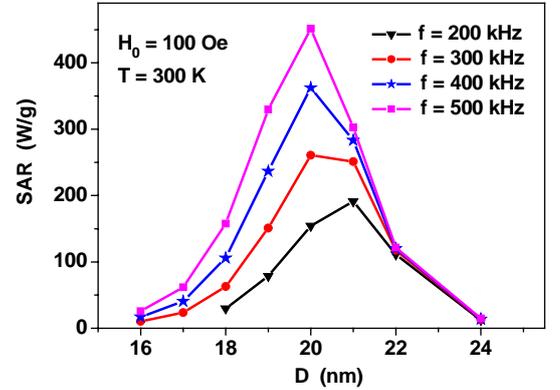

Fig. 2. The specific absorption rate of non-interacting assembly of iron oxides nanoparticles. obtained by means of Eqs. (1) and (2), as a function of average particle diameter at different frequencies of the alternating magnetic field.

However, the experimentally measured SAR values for assemblies of iron oxides nanoparticles are, as a rule, significantly below [18-21] these theoretical values. As we shall see in the next section, this fact can be explained [22-38] by the influence of strong magneto-dipole interaction in dense assemblies of magnetic nanoparticles.

**Assembly of 3D clusters**

Consider now the hysteresis loops of a dilute assembly of 3D random clusters, having easy anisotropy axes of individual nanoparticles randomly oriented in space. As Fig. 2 shows, for the assembly of non-interacting iron oxide nanoparticles the peak of the energy absorption in alternating magnetic field corresponds to particles with diameter $D = 20$ nm. Therefore, first we calculated the hysteresis loops of an assembly of 3D clusters with particle diameter $D = 20$ nm.

Fig. 3a shows the evolution of the assembly hysteresis loops depending on the average distance between the nanoparticle centers $D_{av}$ at the fixed value of the particle diameter $D$. The frequency and amplitude of alternating magnetic field are fixed at $f = 400$ kHz, and $H_0 = 100$ Oe, respectively. The number of particles in the clusters equals $N_p = 40$. The calculations are carried out at $T = 300$ K, magnetic damping constant is taken to be $\kappa = 0.5$.

Evidently, the decrease of the average distance between the nanoparticles of the cluster leads to an increase of intensity of the magneto-dipole interaction within the cluster. Note that for $N_p = 40$ the ratios $D_{av}/D$ specified in Fig. 3a correspond to the cluster packing densities $\eta = 0.005$, 0.04 and 0.32, correspondingly. One can see in Fig. 3a that the hysteresis loop area rapidly decreases as a function of the parameter $\eta$. For comparison, Fig. 3a also shows the hysteresis loop 4), calculated for an assembly of non-interacting particles, i.e. in the limit $D_{av}/D \to \infty$, $N_p = $ const, using the Eqs. (1) and (2).

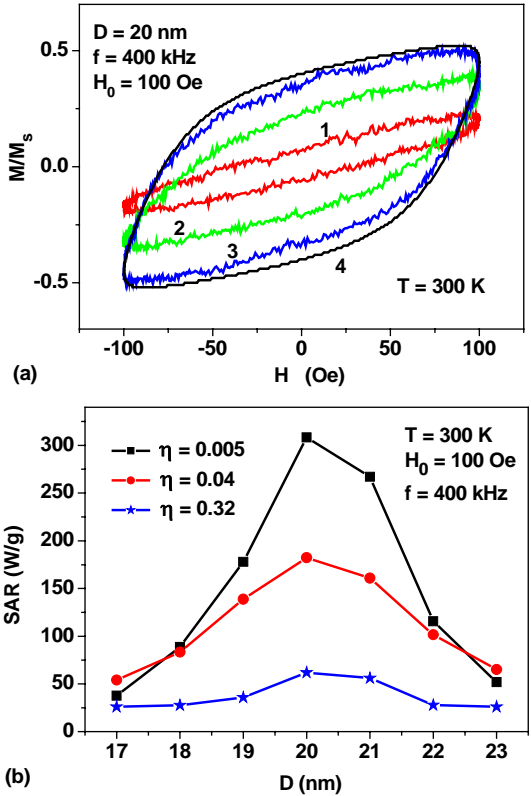

Fig. 3. a) Evolution of the hysteresis loops of dilute assembly of clusters of iron oxide nanoparticles with diameter $D = 20$ nm for various ratios $D_{av}/D$: 1) $D_{av}/D = 1.46$; 2) $D_{av}/D = 2.92$; 3) $D_{av}/D = 5.84$. Hysteresis loop 4) corresponds to assembly of non-interacting nanoparticles of the same diameter. b) SAR as a function of the average nanoparticle diameter $D$ for dilute assemblies of clusters of nanoparticles with different packing density $\eta$.

One can see that the hysteresis loop 3) ($\eta = 0.005$) in Fig. 3a turns out to be close to the hysteresis loop of the assembly of non-interacting nanoparticles. Therefore, in the case $\eta \leq 0.005$ the magneto-dipole interaction of the nanoparticles within the cluster can be neglected. However, for $\eta \geq 0.04$ the magneto-dipole interaction has a significant influence on the properties of an assembly of random 3D clusters. A similar evolution of the assembly hysteresis loops has been obtained also for the frequencies $f = 300$ and 500 kHz, respectively.

The hysteresis loops shown in Fig. 3a are calculated for different ratios $D_{av}/D$, but for the fixed number of nanoparticles in the cluster $N_p = 40$. However, the detailed computer simulations show that the shape of the hysteresis loop of a dilute assembly of random 3D clusters is practically unchanged, if the number of particles, $N_p \gg 1$, and the radius of the cluster $R_{cl}$ are changed so that the nanoparticle packing density $\eta$ remains constant. Therefore, the hysteresis loop of dilute assembly of random 3D clusters depends mainly on the cluster packing density $\eta$.

Fig. 3b shows the SAR of assemblies of random clusters of iron oxide nanoparticles for different $\eta$ values. The SAR of the assembly is calculated [10] as $SAR = 10^{-7} M_s f A / \rho$ (W/g), where $A$ is the hysteresis loop area in the variables ($M/M_s$, $H$), $\rho$ being the density of iron oxide nanoparticles which is assumed to be $\rho = 5$ g/cm$^3$. As Fig. 3b shows, the SAR decreases as a function of $\eta$ due to increase of the intensity of magneto-dipole interaction within the clusters. At the same time, the dependence of the assembly SAR on the average particle diameter still remains, though it becomes less pronounced.

For small values of $\eta \leq 0.005$ the SAR of random assembly of 3D clusters actually coincides with that one for an assembly of non-interacting nanoparticles, shown in Fig. 2. On the other hand, SAR falls about 6 times when the cluster packing density increases up to $\eta = 0.32$. Then it becomes close to typical SAR values $\sim 50 - 100$ W/g, which are obtained in a number of experiments [5,18-21] with iron oxide nanoparticle assemblies.

**Assembly of fractal clusters**

Similar calculations were carried out for dilute assemblies of fractal clusters of nanoparticles with various fractal descriptors. As Fig. 4 shows, for fractal clusters of nanoparticles the SAR as a function of the particle diameter also decreases considerably with respect to that of the assembly of non interacting nanoparticles. However, in contrast to the assembly of 3D clusters the peak values of SAR are shifted systematically to smaller particle diameters, except for the case of fractal dimension $D_f = 2.7$, which is close to the case of 3D clusters with $D_f = 3.0$. It is interesting to note also that for non optimal nanoparticle diameters, for example for nanoparticles with diameters $D \leq 17$ nm, the influence of magneto-dipole interaction leads to increase of the SAR with respect to the case of assembly of non interacting nanoparticles, as the SAR of the assembly of non interacting nanoparticles is very small for nanoparticles with diameters $D \leq 17$ nm.

The calculations shown in Fig.4 were carried out assuming the existence of thin non-magnetic shells with thickness $t_{Sh} = 1$ nm at the surface of magnetic nanoparticles. This prevents the nanoparticles of the fractal cluster from direct exchange interaction. Evidently, the increase of the non-magnetic shell thickness reduces the intensity of the magneto- dipole interaction of closest nanoparticles, as the average distance between the magnetic cores on the nanoparticles increases.

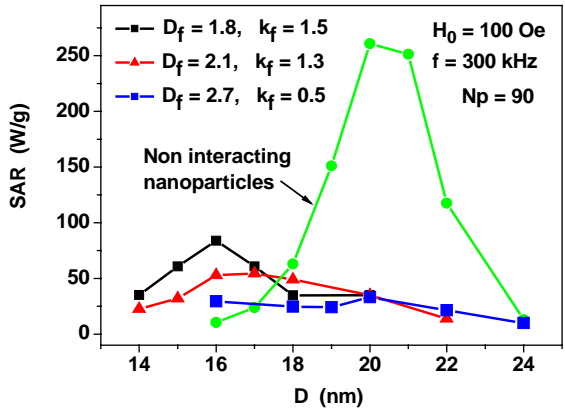

Fig. 4. SAR as a function of the average nanoparticle diameter $D$ for dilute assemblies of fractal clusters of nanoparticles with various fractal descriptors. The SAR of the assembly of non interacting nanoparticles is calculated by means of Eqs. (1), (2).

Fig. 5 shows that the increase of the non-magnetic shell thickness is a proper way to raise the SAR of the assembly of fractal clusters of nanoparticles. Namely, for sufficiently large thickness of non magnetic shells the dependence of the SAR on the particle diameter resembles that for weakly interacting magnetic nanoparticles. This fact may be important for application of magnetic nanoparticle assemblies in magnetic hyperthermia.

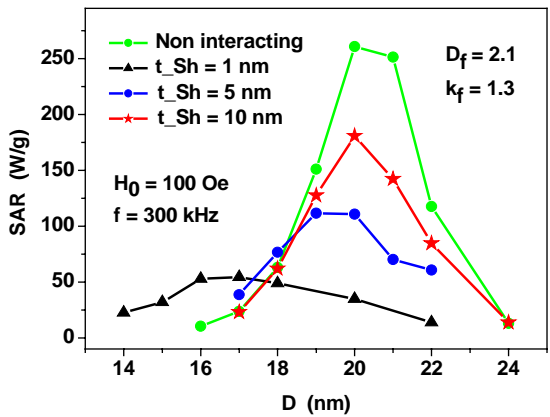

Fig. 5. The dependence of the SAR of dilute assembly of fractal clusters on the thickness $t_{Sh}$ of the non-magnetic shells at the surface of the nanoparticles. The SAR of the assembly of non interacting nanoparticles is calculated by means of Eqs. (1), (2).

**Conclusions**

The main conclusion of this study is that the SAR of a dilute assembly of clusters of magnetic nanoparticles in alternating magnetic field is significantly reduced with increasing of the intensity of magneto-dipole interaction in the clusters. For usual 3D clusters of nanoparticles the intensity of the magneto-dipole interaction can be characterized by dimensionless packing density, $\eta = N_p V/V_{cl} = (D/D_{av})^3$. The latter determines the average distance between the nanoparticles of the cluster. The calculations show that for assembly of random 3D clusters the energy absorption peak, which for iron oxide nanoparticles corresponds to particles with average diameter $D = 20$ nm, is reduced about 6 times when the packing density increases from $\eta = 0.005$ up to $\eta = 0.32$. The dependence of the assembly SAR on the mean nanoparticle diameter is retained with increase of $\eta$, but becomes less pronounced.

For dilute assemblies of fractal clusters of magnetic nanoparticles the SAR values also decrease several times irrespective on the fractal descriptors of the assembly. In addition, the peak values of SAR are shifted systematically to smaller particle diameters, as a rule. It is important to note however, that the increase of the non magnetic shell thickness at the nanoparticle surfaces restore the SAR values close to that of the assembly of weakly interacting nanoparticles. This fact can be important for various biomedical applications of magnetic nanoparticle assemblies.

The model considered in this paper takes into account the geometrical structure of nanoparticle assemblies observed experimentally in biological media [4,8,9] (in particular in tumors), i.e. the agglomeration of nanoparticles in a sufficiently dense fractal clusters of different sizes, with different numbers of nanoparticles in the clusters. The stochastic LL equation (3) accurately describes the real dynamics of the magnetic moments of nanoparticles taking into account both the magneto-dipole interaction between the particles and the effect of thermal fluctuations. The cluster model studied allows obvious generalization that can make it more practical. First, it is necessary to take into account the size distribution of magnetic nanoparticles in the assembly. Second, in some cases exchange interaction may exist between neighboring nanoparticles of the cluster if they are in direct atomic contact.

The theoretical results obtained in this study seem to be in a satisfactory agreement with recent experimental data [35] for iron oxide nanoparticles of optimal diameters. Indeed, according to Ref. 35 the SAR of the iron oxide nanoparticles increases with the average diameter of the nanoparticles and peaks for nanoparticles with mean diameter $D = 20 - 21$ nm. In addition, the SAR decreases [35] with a decrease in the average distance between the nanoparticles due to increasing intensity of the magneto-dipole interaction.

Unfortunately, in some experimental studies [5,21] carried out to optimize the properties of magnetic nanoparticles for use in magnetic hyperthermia, often do not take into account the theoretical predictions [10,11] about significant dependence of the assembly SAR on the characteristic size of the magnetic nanoparticles. As shown in this paper, this dependence can be substantial even for rather dense nanoparticle assemblies. From a theoretical point of view, it is obvious [10] that the assembly of iron oxide nanoparticles with very small, $D \leq 10$ nm, or too big, $D \geq 30$ nm diameters can hardly provide a sufficiently high SAR values for typical for magnetic hyperthermia frequencies, $f = 200 - 600$ kHz, and magnetic field amplitudes $H_0 \sim 100$ Oe. The creation of mono- crystalline iron oxide nanoparticles with sharp size distribution near optimal diameter has to be promising for application in magnetic hyperthermia.

**Acknowledgement**

The authors gratefully acknowledge the financial support of the Ministry of Education and Science of the Russian Federation in the framework of Increase Competitiveness Program of NUST «MISIS», contract № K2-2015-018.

**References**

1. Pankhurst QA, Thanh NKT, Jones SK, Dobson J (2009) Progress in applications of magnetic nanoparticles in biomedicine. J Phys D: Appl Phys 42:224001.
2. Dutz S, Hergt R (2013) Magnetic nanoparticle heating and heat transfer on a microscale: Basic principles, realities and physical limitations of hyperthermia for tumour therapy. Int J Hyperthermia 29**:**790-800.
3. Ortega D, Pankhurst QA Magnetic hyperthermia. In O'Brien P editor. Nanoscience: Nanostructures through Chemistry. Royal Society of Chemistry: Cambridge 2013 1:60–88.
4. Périgo EA, Hemery G, Sandre O, Ortega D, Garaio E, Plazaola F, Teran FJ (2015) Fundamentals and advances in magnetic hyperthermia. Appl Phys Rev 2:041302.
5. Cervadoro A, Giverso C, Pande R, Sarangi S, Preziosi L, Wosik J, Brazdeikis A, Decuzzi P (2013) Design Maps for the Hyperthermic Treatment of Tumors with Superparamagnetic Nanoparticles. PLoS One 8**:**e57332.
6. Dutz S, Kettering M, Hilger I, Muller R, Zeisberger M (2011) Magnetic multicore nanoparticles for hyperthermia – influence of particle immobilization in tumour tissue on magnetic properties. Nanotechnology 22:265102.
7. Dutz S, Hergt R (2012) The role of interactions in systems of single domain ferrimagnetic iron oxide nanoparticles. J Nano- Electron Phys 4:20101-20107.
8. Etheridge ML, Hurley KR, Zhang J, Jeon S, Ring HL, Hogan C, Haynes CL, Garwood M, Bischof JC (2014) Accounting for biological aggregation in heating and imaging of magnetic nanoparticles. Technology 2:214-228.
9. Jeon S, Hurley KR, Bischof JC, Haynes CL, Hogan CJ, Jr (2016) Quantifying intra-and extracellular aggregation of iron oxide nanoparticles and its influence on specific absorption rate. Nanoscale 8:16053-16064.
10. Usov NA (2010) Low frequency hysteresis loops of superparamagnetic nanoparticles with uniaxial anisotropy. J Appl Phys 107**:**123909.
11. Carrey J, Mehdaoui B, Respaud M (2011) Simple models for dynamic hysteresis loop calculations of magnetic single-domain nanoparticles: Application to magnetic hyperthermia optimization. J Appl Phys 109:083921.
12. Mamiya H, Jeyadevan B (2011) Hyperthermic effects of dissipative structures of magnetic nanoparticles in large alternating magnetic fields. Sci Rep 1:157.
13. Usov NA, Liubimov BYa (2012) Dynamics of magnetic nanoparticle in a viscous liquid: application to magnetic nanoparticle hyperthermia. J Appl Phys 112:023901.
14. Hergt R, Hiergeist R, Zeisberger M, Schüller D, Heyen U, Hilger I, Kaiser WA (2005) Magnetic properties of bacterial magnetosomes as potential diagnostic and therapeutic tools. J Magn Magn Mater 293:80-86.
15. Zeisberger M, Dutz S, Müller R, Herdt R, Matoussevitch N, Bönnemann H (2007) Metallic cobalt nanoparticles for heating applications J Magn Magn Mater 311:224-233.
16. Mehdaoui B, Meffre A, Lacroix L-M, Carrey J, Lachaize S, Gougeon M, Respaud M, Chaudret B (2010) Large specific absorption rates in the magnetic hyperthermia properties of metallic iron nanocubes. J Magn Magn Mater 322:L49-L52.
17. Mehdaoui B, Meffre A, Carrey J, Lachaize S, Lacroix L-M, Gougeon M, Chaudret B, Respaud M (2011) Optimal size of nanoparticles for magnetic hyperthermia: a combined theoretical and experimental study. Adv Funct Mater 21:4573-4581.
18. Kallumadil M, Tada M, Nakagawa T, Abe M, Southern P, Pankhurst QA (2009) Suitability of commercial colloids for magnetic hyperthermia. J Magn Magn Mater 321:1509-1513.
19. Li Z, Kawashita M, Araki N, Mitsumori M, Hiraoka M et al. (2010) Magnetite nanoparticles with high heating efficiencies for application in the hyperthermia of cancer. Materials Science and Engineering: C 30:990–996.
20. Li CH, Hodgins P, Peterson GP (2011) Experimental study of fundamental mechanisms in inductive heating of ferromagnetic nanoparticles suspension (Fe3O4 Iron Oxide Ferrofluid. J Appl Phys 110:054303.
21. Smolkova IS, Kazantseva NE, Babayan V, Smolka P, Parmar H, Vilcakova J, Schneeweiss O, Pizurova N (2015) Alternating magnetic field energy absorption in the dispersion of iron oxide nanoparticles in a viscous medium. J Magn Magn Mater 374:508-515.
22. Gudoshnikov SA, Liubimov BYa, Usov NA( 2012) Hysteresis losses in a dense superparamagnetic nanoparticle assembly. AIP Advances 2:012143.
23. Gudoshnikov SA, Liubimov BYa, Popova AV, Usov NA (2012) The influence of a demagnetizing field on hysteresis losses in a dense assembly of superparamagnetic nanoparticles. J Magn Magn Mater 324:3690-3695.
24. Dennis CL, Jackson AJ, Borchers JA, Hoopes PJ, Strawbridge R, Foreman AR, Van Lierop J, Grutner C, Ivkov R (2009) Nearly complete regression of tumors via collective behavior of magnetic nanoparticles in hyperthermia. Nanotechnology 20:395103.

25. Serantes D, Baldomir D, Martinez-Boubeta C, Simeonidis K, Angelakeris M, Natividad E et al. (2010) Influence of dipolar interactions on hyperthermia properties of ferromagnetic particles. J Appl Phys 108:073918.
26. Urtizberea A, Natividad E, Arizaga A, Castro M, Mediano A (2010) Specific absorption rates and magnetic properties of ferrofluids with interaction effects at low concentrations. J Phys Chem C 114:4916-4922.
27. Burrows F, Parker C, Evans RFL, Hancock Y, Hovorka O, Chantrell RW (2010) Energy losses in interacting fine-particle magnetic composites. J Phys D: Appl Phys 43:474010.
28. Eberbeck D, Trahms L (2011) Experimental investigation of dipolar interaction in suspensions of magnetic nanoparticles. J Magn Magn Mater 323:1228-1232.
29. Martinez-Boubeta C, Simeonidis K, Serantes D, Leboran IC, Kazakis I, Sefanou G, Pena L, Galceran R, Balcells L, Monty C, Baldomir D, Mitrakas M, Angelakeris M (2012) Adjustable Hyperthermia Response of Self-Assembled Ferromagnetic Fe-MgO Core–Shell Nanoparticles by Tuning Dipole–Dipole Interactions. Adv Funct Mater 22:3737-3744.
30. Branquinho LC, Carriao MS, Costa AS, Zufelato N, Sousa MH, Miotto R, Ivkov R, Bakuzis AF (2013) Effect of magnetic dipolar interactions on nanoparticle heating efficiency: Implications for cancer hyperthermia. Sci. Rep 3:2887.
31. Mehdaoui B, Tan RP, Meffre A, Carrey J, Lachaize S, Chaudret B, Respaud M (2013) Increase of magnetic hyperthermia efficiency due to dipolar interactions in low-anisotropy magnetic nanoparticles: Theoretical and experimental results. Phys Rev B 87:174419.
32. Serantes D, Simeonidis K, Angelakeris M, Chubykalo-Fesenko O, Marciello M, Del Puerto Morales M, Baldomir D, Martinez-Boubeta C (2014) Multiplying magnetic hyperthermia response by nanoparticle assembling. J Phys Chem C 118:5927-5934.
33. Landi GT (2014) Role of dipolar interaction in magnetic hyperthermia. Phys Rev B 89:014403.
34. Tan RP, Carrey J, Respaud M (2014) Magnetic hyperthermia properties of nanoparticles inside lysosomes using kinetic Monte Carlo simulations: Influence of key parameters and dipolar interactions, and evidence for strong spatial variation of heating power. Phys Rev B 90:214421.
35. Blanco-Andujar C, Ortega D, Southern P, Pankhurst QA, Thanh NTK (2015) High performance multi-core iron oxide nanoparticles for magnetic hyperthermia: microwave synthesis, and the role of core-to-core interactions. Nanoscale 7:1768-1775.
36. Conde-Leboran I, Baldomir D, Martinez-Boubeta C, Chubykalo-Fesenko O, Morales MD, Salas G, Cabrera D, Camarero J, Teran FJ, Serantes D (2015) A single picture explains diversity of hyperthermia response of magnetic nanoparticles. J Phys Chem C 119:15698-15706.
37. Ruta S, Chantrell R, Hovorka O (2015) Unified model of hyperthermia via hysteresis heating in systems of interacting magnetic nanoparticles. Sci Rep 5:9090.
38. Sanz B, Calatayud MP, Biasi ED, Lima E Jr., Mansilla MV, Zysler RD, Ibarra MR, Goya GF (2016) In silico before in vivo: how to predict the heating efficiency of magnetic nanoparticles within the intracellular space. Sci Rep 6:38733.
39. Brown Jr. WF (1963) Thermal fluctuations of a single-domain particle. Phys Rev 130:1677-1686.
40. Garcia-Palacios JL, Lazaro F (1998) Langevin-dynamics study of the dynamical properties of small magnetic particles. J Phys Rev B 58:14937-14958.
41. Scholz W, Schrefl T, Fidler J (2001) Micromagnetic simulation of thermally activated switching in fine particles. J Magn Magn Mater 233:296-304.
42. Forrest SR, Witten TA Jr. (1979) Long-range correlations in smoke-particle aggregates. J Phys A: Math Gen 12:L109-L117.
43. Filippov AV, Zurita M, Rosner DE (2000) Fractal-like aggregates: relation between morphology and physical properties. J Colloid Interface Sci 229:261–273.